\documentstyle[12pt]{article}
\begin{document}
\title{\bf Some remarks on the interpretation of degree of nonextensivity}
\author{G.Wilk$^{1}$\thanks{e-mail: wilk@fuw.edu.pl} and Z.W\l
odarczyk$^{2}$ \thanks{e-mail: wlod@pu.kielce.pl}\\[2ex] 
$^1${\it The Andrzej So\l tan Institute for Nuclear Studies}\\
    {\it Ho\.za 69; 00-689 Warsaw, Poland}\\
$^2${\it Institute of Physics, Pedagogical University}\\
    {\it  Konopnickiej 15; 25-405 Kielce, Poland}}  
\date{\today}
\maketitle

\begin{abstract}
Recently we have demostrated that the nonextensitivity parameter $q$
occuring in some applications of Tsallis statistics (known also as
index of the corresponding L\'evy  distribution) is, in $q>1$ case,
given entirely by the fluctuations of the parameters of the usual 
exponential distribution. We show here that this interpretation is
valid also for the $q<1$ case. The parameter $q$ is therefore a
measure of fluctuations of the parameters of the usual exponential
distribution.\\   

[3ex]

\end{abstract}

\newpage

Recently we have demostrated that the nonextensitivity parameter $q$
occuring in some applications of Tsallis statistics \cite{T} (known
also as index of the corresponding L\'evy  distribution) is, in $q>1$
case, given entirely by the fluctuations of the parameters of the usual 
exponential distribution \cite{WWq}. It means that:
\begin{itemize}
\item when in some exponential formula describing distribution of a
quantity $\varepsilon$ of physical interest:
\begin{equation}
L_{q=1}(\varepsilon)\, =\, C_{q=1}\, \exp\left[ -
                    \frac{\varepsilon}{\chi}\right] , \label{eq:Lq1}
\end{equation}
one allows the parameter $\chi$ to fluctuate around some mean value
$\chi_0$, and 
\item if these fluctuations are described by simple Gamma
distribution of the form  
\begin{equation}
f(\frac{1}{\chi})\, =\, \frac{1}{\Gamma(\alpha)}\, \mu\, 
 \left(\frac{\mu}{\chi}\right)^{\alpha-1}\, \exp\left( -\,
\frac{\mu}{\chi} \right)  \label{eq:FRES}
\end{equation}
depending on two parameters
\begin{equation}
\alpha \, =\, \frac{1}{q-1}\qquad {\rm and} \qquad \mu = \alpha
               \chi_0 , \label{eq:param}
\end{equation}
\item then, as result, one gets the following power-like distribution
for the quantity $\varepsilon$ of interest:
\begin{equation}
L_q(\varepsilon)\, =\, C_q\, \left[ 1\, -\, (1\, -\, q)\, 
                    \frac{\varepsilon}{\chi_0}\right]^{\frac{1}{1-q}},
                    \label{eq:Lq}
\end{equation}                    
known also as L\'evy distribution with index $q$, where
\begin{equation}
q\, =\, 1\, +\, 
       \frac{\left\langle\left(\frac{1}{\chi}\right)^2\right\rangle\,
       -\, \left\langle\frac{1}{\chi}\right\rangle^2}
       {\left\langle \frac{1}{\chi}\right\rangle^2}, \label{eq:q}
\end{equation}
i.e., where it is entirely given by the relative variance of  the
parameter $1/\chi$ of the initial
distribution (\ref{eq:Lq1}) ($<...>$ denotes the corresponding
averages with respect to distribution $f(\chi)$). 
\end{itemize}
The proof presented in \cite{WWq} was limited to the $q>1$ case and
the physical discussion provided there was also concentrated on such
situation. Because $q$ can be also interpreted as the so called
nonextensivity parameter occuring in some applications of Tsallis 
statistic \cite{T}, it would be interesting to check if such
interpretation can be extended to the $q<1$ case as well. We shall
demonstrate below that this is indeed the case \cite{FOOT1}.\\

The essential difference between these two cases is, for the purpose
of present discussion, that whereas for $q>1$ probability distribution
$L_q(\varepsilon)$ is well defined for the whole range of variable
$\varepsilon$, $\varepsilon \in (0,\infty)$, for $q<1$ it is defined
only for $\varepsilon \in [0,\chi_0/(1-q)]$. As it was done in
\cite{WWq} we shall deduce the form of function $f(1/\chi)$, describing
fluctuations in $\chi$, which would lead from the exponential
distribution $L_{q=1}$ to the power-like L\'evy distribution $L_{q<1}$
\begin{equation}
L_{q<1}(\varepsilon;\chi_0)\, =\, C_q\, \left[1\, -\,
       \frac{\varepsilon}{\alpha' \chi_0}\right]^{\alpha'}\, =\,
       C_q\, \int^{\infty}_0\, \exp\left(- \frac{\varepsilon}{\chi}\right)
       \, f\left(\frac{1}{\chi}\right)\, d\left(\frac{1}{\chi}\right)
       \label{eq:QL1}
\end{equation}       
(for simplicity we denote $\alpha' = \frac{1}{1-q}$). From the
representation of the Euler gamma function we have
\begin{equation}
\left[1\, -\, \frac{\varepsilon}{\alpha' \chi_0}\right]^{\alpha'}\, =\,
\left(\frac{\alpha' \chi_0}{\alpha' \chi_0 - \varepsilon}\right)^{-\alpha'}\,
=\, \frac{1}{\Gamma(\alpha')}\, \int^{\infty}_0\, d\eta\, 
    \eta^{\alpha' - 1}\, \exp\left[ - \eta\, \left(1\, +\, 
    \frac{\varepsilon}{\alpha' \chi_0 - \varepsilon}\right)\right] . 
    \label{eq:EGF}
\end{equation}
Changing now variables under the integral in such a way that
$\chi\, =\, \frac{\alpha' \chi_0 - \varepsilon}{\eta}$
one immediately obtains Eq. (\ref{eq:QL1}) with $f(1/\chi)$ given by
the following gamma distribution
\begin{equation}
f\left(\frac{1}{\chi}\right)\, =\, \frac{1}{\Gamma(\alpha')}\,
     \left( \alpha' \chi_0 - \varepsilon \right)\,
     \left(\frac{\alpha' \chi_0 - \varepsilon}{\chi}\right)^{\alpha' - 1}\,
     \exp\left( - \frac{\alpha' \chi_0 - \varepsilon}{\chi}\right) 
     \label{eq:RES}
\end{equation}
with parameters $\alpha'$ and $\mu(\varepsilon) = \alpha' \chi_0 -
\varepsilon$. This time, contrary to the $q>1$ case of \cite{WWq},
fluctuations depend on the value of the variable in question, i.e.,
the mean value and variance are both $\varepsilon$-dependent:
\begin{equation}
\left\langle \frac{1}{\chi}\right\rangle\, =\, \frac{1}{\chi_0 -
\frac{\varepsilon}{\alpha'}}\qquad {\rm and}\qquad \left\langle
\left(\frac{1}{\chi}\right)^2\right\rangle\, -\,
\left\langle\frac{1}{\chi}\right\rangle^2\, =\, \frac{1}{\alpha'}\cdot
\frac{1}{\left(\chi_0 - \frac{\varepsilon}{\alpha'}\right)^2} .
\label{eq:MV}
\end{equation}
However, the relative variance
\begin{equation}
\omega\, =\,  \frac{\left\langle\left(\frac{1}{\chi}\right)^2\right\rangle\,
       -\, \left\langle\frac{1}{\chi}\right\rangle^2}
       {\left\langle \frac{1}{\chi}\right\rangle^2}\, 
       =\, \frac{1}{\alpha'}\, =\, 1\, -\, q \label{eq:RESULT}
\end{equation}
remains $\varepsilon$-independent (exactly like in the case of $q>1$)
and depends only on parameter $q$. It means therefore that the
parameter $q$ in L\'evy distribution $L_q(\varepsilon)$ describes the
relative variance of fluctuations of parameter $\chi$ in
$L_{q=1}(\varepsilon)$ for all values of $q$ (both for $q>1$, where
$\omega = q - 1$, cf. \cite{WWq} and for $q<1$ as given above, where
$\omega = 1-q$).\\ 

In \cite{WWq} we have proposed a general explanation of the meaning
of function $f(\chi)$ describing fluctuations of some variable $\chi$.
The question one is interested in is why, and under what
circumstances, it is the gamma distribution that describes 
fluctuations. To this end we have started with general Langevin type
equation \cite{FP} for the variable $\chi$ 
\begin{equation}
\frac{d\chi}{dt}\, +\, \left[\frac{1}{\tau}\, +\, \xi(t)\right]\,
\chi\, =\, \phi\, =\, {\rm const}\, >\, 0  \label{eq:LE}
\end{equation}
(with damping constant $\tau$ and source term $\phi$). For stochastic
processes defined by the white gaussian noise form of $\xi(t)$ (cf.
\cite{WWq} for details) it can be shown that distribution function
for the variable $\chi$ satisfies the Fokker-Planck equation
($K_{1,2}$ are the corresponding intensity coefficients, cf.
\cite{WWq}) 
\begin{equation}
\frac{df(\chi)}{dt}\, =\, -\, \frac{\partial}{\partial \chi}K_1\,
f(\chi)\, +\, \frac{1}{2}\, \frac{\partial^2}{\partial \chi^2}K_2\,
f(\chi) , \label{eq:FPE}
\end{equation}
i.e., it is indeed given by the Gamma distribution in variable
$1/\chi$ of the form (\ref{eq:FRES}) with $\mu = \alpha\chi_0$.
Notice that it differs from Eq. (\ref{eq:RES}) only in the form of
parameter $\mu$, which in (\ref{eq:RES}) depends also on the physical
quantity of interest $\varepsilon$.\\

As an illustration of the genesis of Eq. (\ref{eq:LE}) we have
discussed in \cite{WWq} the case of fluctuations of temperature
(i.e., the situation where $\chi=T$) \cite{FOOT2}. Suppose that we
have a thermodynamic system, in a small (mentally separated) part of
which the temperature fluctuates around some mean value $T_0$ (which
can be also understood as an equilibrium temperature) with
$\Delta T \sim T$. The unevitable exchange of heat between this
selected region and the rest of the system is described by Eq.
(\ref{eq:LE}) in which 
\begin{equation}
\phi = \phi_{q<1}\, =\, \frac{1}{\tau}\left(T_0 -
                 \frac{\varepsilon}{\alpha'}\right) 
\qquad {\rm whereas}\qquad \phi = \phi_{q>1} = \frac{T_0}{\tau} .
             \label{eq:FIFI}
\end{equation}
It means that the corresponding process of heat conductivity is, for
$q<1$ case, described by the following equation (here $T'=T_0-\tau
\xi(t)T$) 
\begin{equation}
\frac{\partial T}{\partial t}\, -\, \frac{1}{\tau}
            \, (T'\, -\, T)\,
         +\, \frac{\varepsilon}{\tau \alpha'} =\, 0 , \label{eq:HC}
\end{equation}
which differs from the corresponding equation for $q>1$ case only by
the last term describing the presence the internal heat source. It
has a sense of dissipative transfer of energy from the region where
(due to fluctuation) we have higher $T$. It could be any kind of
convection type flow of energy, for example it could be connected
with emission of particles from that region. The heat release given
by $\varepsilon/(\tau\alpha')$ depends on $\varepsilon$ (but it is
only a part of $\varepsilon$, which is released). In the case of such
energy release (connected with emission of particles) there is
additionale cooling of the whole system. If this process is
sufficiently fast, it could happen that there is no way to reach a
stationary distribution of temperature (because the transfer of the
heat from the outside can be not sufficient for development of the
state of equilibrium). On the other hand (albeit this is not our case
here) for the reverse process we could face the "heat explosion"
situation (which could happen if the velocity of the exotermic
burning reaction grows sufficiently fast; in this case because of
nonexistence of stationary distribution we have fast nonstationary
heating of the substance and acceleration of the respective reaction).\\   

It should ne noticed that in the case of $q<1$ the temperature does
not reach stationary state because, cf. Eq. (\ref{eq:MV}),
$\langle 1/T \rangle\, =\, 1/(T_0 - \varepsilon/\alpha')$,
whereas for $q>1$ we had $<T> = T_0$. As a consequence the
corresponding L\'evy distribution are defined only for $\varepsilon
\in(0, T_0\, \alpha'$) because for $\varepsilon \rightarrow
T_0\alpha'$ the $<T>\rightarrow 0$. Such asymptotic (i.e., for
$t/\tau \rightarrow \infty$) cooling of the system ($T\rightarrow 0$)
can be also deduced form Eq. (\ref{eq:HC}) for $\varepsilon \rightarrow
T_0\alpha'$.\\

To summarize, we have demonstrated that temperature fluctuations lead
to the L\'evy distribution $L_q(\varepsilon)$ with index $q<1$ when
there exists energy source and with $q>1$ in the absence of such
source. In both cases, however, the relative variance of $1/T$ 
fluctuations is described by the parameter $q$ only.\\


\begin{thebibliography}{99}
\bibitem{FOOT1} The relevant references to the up-to-date applications
                 of the Tsallis statistics can be found in \cite{T}
                 whereas \cite{WWq} summarises all recent attempts of
                 its applications in the high energy physics domain.
\end{thebibliography}
\end{document}